\documentclass[preprint,aps,tightenlines]{revtex4}
\usepackage[dvipdf]{graphicx}

\begin{document}
\title{ Vibration induced phase noise in Mach-Zehnder atom interferometers}

\author{ A. Miffre (1, 2), M. Jacquey (1), M. B\"uchner(1), G. Tr\'enec (1) and J. Vigu\'e(1)}
\address{ Laboratoire Collisions Agr\'egats R\'eactivit\'e -IRSAMC
\\ (1) Universit\'e Paul Sabatier and CNRS UMR 5589
\\118, Route de Narbonne 31062 Toulouse Cedex, France
\\ (2) PIIM, Universit\'e de Provence and CNRS UMR 6633,
\\ Centre de saint J\'er\^ome case C21, 13397 Marseille cedex 20,
France
\\ e-mail:~{\tt jacques.vigue@irsamc.ups-tlse.fr}}

\date{\today}

\begin{abstract}

The high inertial sensitivity of atom interferometers has been
used to build accelerometers and gyrometers but this sensitivity
makes these interferometers very sensitive to the laboratory
seismic noise. This seismic noise induces a phase noise which is
large enough to reduce the fringe visibility in many cases. We
develop here a model calculation of this phase noise in the case
of Mach-Zehnder atom interferometers and we apply this model to
our thermal lithium interferometer. We are thus able to explain
the observed dependence of the fringe visibility with the
diffraction order. The dynamical model developed in the present
paper should be very useful to further reduce this phase noise in
atom interferometers and this reduction should open the way to
improved interferometers.

\end{abstract}

\maketitle

%%%%%%%%%%%%%%%%%%%%%%%%%%%%%%%%%%%%%%%%%%%%%%%%%%%%%%%%%%%%%%%%%%%%%%

\section{Introduction}

Atom interferometers have a large inertial sensitivity
\cite{anandan77,clauser88}, which has been used to develop
sensitive accelerometers \cite{ kasevich91, kasevich92, cahn97,
peters99, peters01, snadden98, mcguirk02,tino02} and gyrometers
\cite{riehle91,lenef97,gustavson97,gustavson00,leduc04,landragin04}.
However, because of this large sensitivity, a high mechanical
stability of the experiment is required. This problem was
recognized in 1991 by D. Pritchard and coworkers \cite{ keith91}
who were obliged to actively control the vibrations of the
diffraction gratings of their Mach-Zehnder thermal atom
interferometer. Since this work, various types of vibration
control were developed: as an example, a very efficient control
was developed by Chu and co-workers \cite{peters99, peters01} for
the measurement of the local acceleration of gravity $g$. The
problem obviously depends on the interferometer design and the
present paper is devoted to an analysis of the vibration problem
in three-grating Mach-Zehnder interferometers operated with
thermal atoms.

In the present paper, we are going to evaluate the phase noise
induced by mechanical vibrations in a Mach-Zehnder thermal atom
interferometer. In our instrument, a very stiff rail holds the
three diffraction gratings and this arrangement has strongly the
effect of vibrations with respect to previous interferometers. We
first analyze how the vibrations displace and distort the rail
holding the gratings, by developing a simple model of the dynamics
of this rail, using elasticity theory. This model will enable us
to understand the contributions of various frequencies and to
prove the importance of the vibration induced rotations of the
rail. The predictions of this model will be tested in the case of
our setup and the phase noise thus evaluated is in good agreement
with the value deduced from fringe visibility measurements.

The paper is organized in the following way : part 2 recalls
classic results concerning the inertial sensitivity of 3-grating
Mach-Zehnder interferometers. Part 3 describes theoretically the
motion and deformation of the rail holding the gratings and the
resulting phase effect. Part 4 describes the rail of our
interferometer and applies the present theory to this case. Part 5
discusses how to further reduce the vibration induced phase noise
in this type of atom interferometers.

\section{Sensitivity of Mach-Zehnder atom interferometers
to accelerations and rotations}

%%%%%%%%%%%%%%%%%%%%%%%%%%%%%%%%%%%%%%%%%%%%%%%%%%%%%%
\begin{figure}
\includegraphics[width = 8 cm,height= 5 cm]{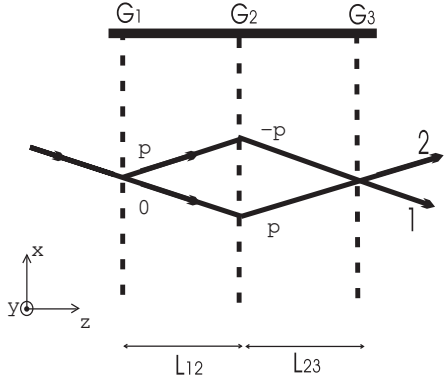}
\caption{\label{mach-zehnder} Schematic drawing of a three grating
Mach-Zehnder atom interferometer, in the Bragg diffraction
geometry. A collimated atomic beam is successively diffracted by
three gratings $G_1$, $G_2$ and $G_3$. The diffraction orders are
indicated on the various paths. Two exit beams, labelled $1$ and
$2$, carry complementary signals. The $x$, $y$, $z$ axis are
defined.}
\end{figure}
%%%%%%%%%%%%%%%%%%%%%%%%%%%%%%%%%%%%%%%%%%%%%%%%%%%%%%

Atom interferometers are very sensitive to inertial effects
\cite{anandan77,clauser88}. We consider a three-grating
Mach-Zehnder atom interferometer represented schematically in
figure \ref{mach-zehnder} and we follow a tutorial argument
presented by Schmiedmayer et al. in reference
\cite{schmiedmayer97}. Each atomic beam is represented by a plane
wave. When a plane wave $\Psi = \exp\left[i{\mathbf k}{\mathbf
r}\right]$ is diffracted by a grating $G_j$, diffraction of order
$p$ produces a plane wave:

\begin{equation}
\label{a1} \Psi_d ({\mathbf r}) =  \alpha_j(p_j)
\exp\left[i{\mathbf k}\cdot{\mathbf r} + i p_j {\mathbf
k}_{Gj}\cdot\left({\mathbf r}- {\mathbf r_j}\right)\right]
\end{equation}

\noindent $\alpha_j(p_j)$ is the diffraction amplitude; $ {\mathbf
k}_{Gj}$ is the grating wavevector, in the grating plane and
perpendicular to its lines, with a modulus $k_{Gj} = 2\pi/a$. The
grating period $a$ is equal to $a= \lambda_L/2$ in the case of
diffraction by a laser standing wave with a laser wavelength
$\lambda_L$. This equation is exact for Bragg diffraction and a
good approximation if ${\mathbf k}$ and ${\mathbf k}_{Gj}$ are
almost perpendicular and $\left|{\mathbf k}_{Gj}\right| \ll
\left|{\mathbf k}\right|$. Finally, ${\mathbf r_j}$ is a
coordinate which measures the position of a reference point in
grating $G_j$. Because of the presence of ${\mathbf r_j}$ in
equation (\ref{a1}), the phase of the diffracted wave depends on
the position of the grating in its plane and this dependence
explains the inertial sensitivity of atom interferometers.

It is easy to calculate the waves exiting from the interferometer
by the exit $1$ in figure \ref{mach-zehnder}, one wave $\Psi_u$
following the upper path with diffraction orders $p$, $-p$ and $0$
and the other wave $\Psi_l$ following the lower path with the
diffraction orders $0$, $p$ and $-p$. These two waves produce an
intensity proportional to $ \left| \Psi_u + \Psi_l \right|^2$,
which must be integrated over the detector surface. The condition
${\mathbf k}_{G1} + {\mathbf k}_{G3} = 2{\mathbf k}_{G2}$ must be
fulfilled to maximize the fringe visibility. We will assume that
this condition is realized and that the grating wavevectors
${\mathbf k}_{Gi} $ are parallel to the $x$-axis. Then, the
interferometer output signal $I$ measured at exit $1$ is given by:

\begin{equation}
\label{a4}
 I =  I_{m} \left[1 + {\mathcal{V}}\cos \Phi_p \right]
 \mbox{   with   } \Phi_p = p k_G
 \left[2 x_2 - x_1 -x_3 \right]
\end{equation}

\noindent where $I_{m}$ is the mean intensity, ${\mathcal{V}}$ is
the fringe visibility defined by $ {\mathcal{V}} =\left(I_{max} -
I_{min}\right)/\left(I_{max} + I_{min}\right)$. When the gratings
are moving, we must correct the grating-position dependent phase
$\Phi$ in equation (\ref{a4}) by considering for each atomic wave
packet the position of the grating $G_j$ at the time $t_j$ when
the wavepacket goes through this grating:

\begin{equation}
\label{a5} \Phi_p = p k_G \left[2 x_2(t_2) - x_1(t_1) -x_3(t_3)
\right]
\end{equation}

\noindent If $=L_{12}/u$ is the atom time of flight $T=L_{12}/u$
from one grating to the next (with $L_{12} = L_{23}$ and $u$ being
the atom velocity), $t_j$ are given by $t_1 = t-T$ and $t_3 =
t+T$, where $t_2$ has been noted $t$. We can expand $\Phi$ in
powers of $T$ by introducing the $x$-components of the velocity
$v_{jx}(t)$ and acceleration $a_{jx}(t)$ of grating $G_j$ measured
with reference to a Galilean frame. The phase $\Phi_p$ becomes:

\begin{equation}
\label{a6} \Phi_p= \Phi_{bending} +\Phi_{Sagnac} + \Phi_{acc.}
\end{equation}

\noindent with $\Phi_{bending} =p  k_G \delta(t)$ where the
bending $\delta(t) = 2 x_2(t) -x_1(t) -x_3(t)$ is so called
because it vanishes when the three gratings are aligned. The
second term represents Sagnac effect because the velocity
difference can be written $ \left( v_{3x} - v_{1x} \right) = 2
\Omega_y L_{12}$, where $\Omega_y$ is the $y$-component of the
angular velocity of the interferometer rail. Finally, the third
term $\Phi_{acc.} =  p  k_G \left(a_{1x} + a_{3x}\right) T^2/2$
describes the sensitivity to linear acceleration \cite{anandan77},
slightly modified because the accelerations of the gratings $G_1$
and $G_3$ are different.

\section{Theoretical analysis of the rail dynamics}

To calculate the phase $\Phi$, we are going to relate the
positions $x_j(t_j)$ of the three gratings to the mechanical
properties of the rail holding the three gratings and to its
coupling to the environment. A 1D theory of the rail is sufficient
to describe the grating motions in the $x$ direction but we want
to know the $x_j(t_j)$ functions typically up to $10^3$ Hertz and
the rail must be treated as an elastic object well before reaching
$10^3$ Hz.

\subsection{Equations of motion of the rail deduced from elasticity theory}

%%%%%%%%%%%%%%%%%%%%%%%%%%%%%%%%%%%%%%%%%%%%%%%%%%%%%%
\begin{figure}
\includegraphics[width = 12 cm,height= 8 cm]{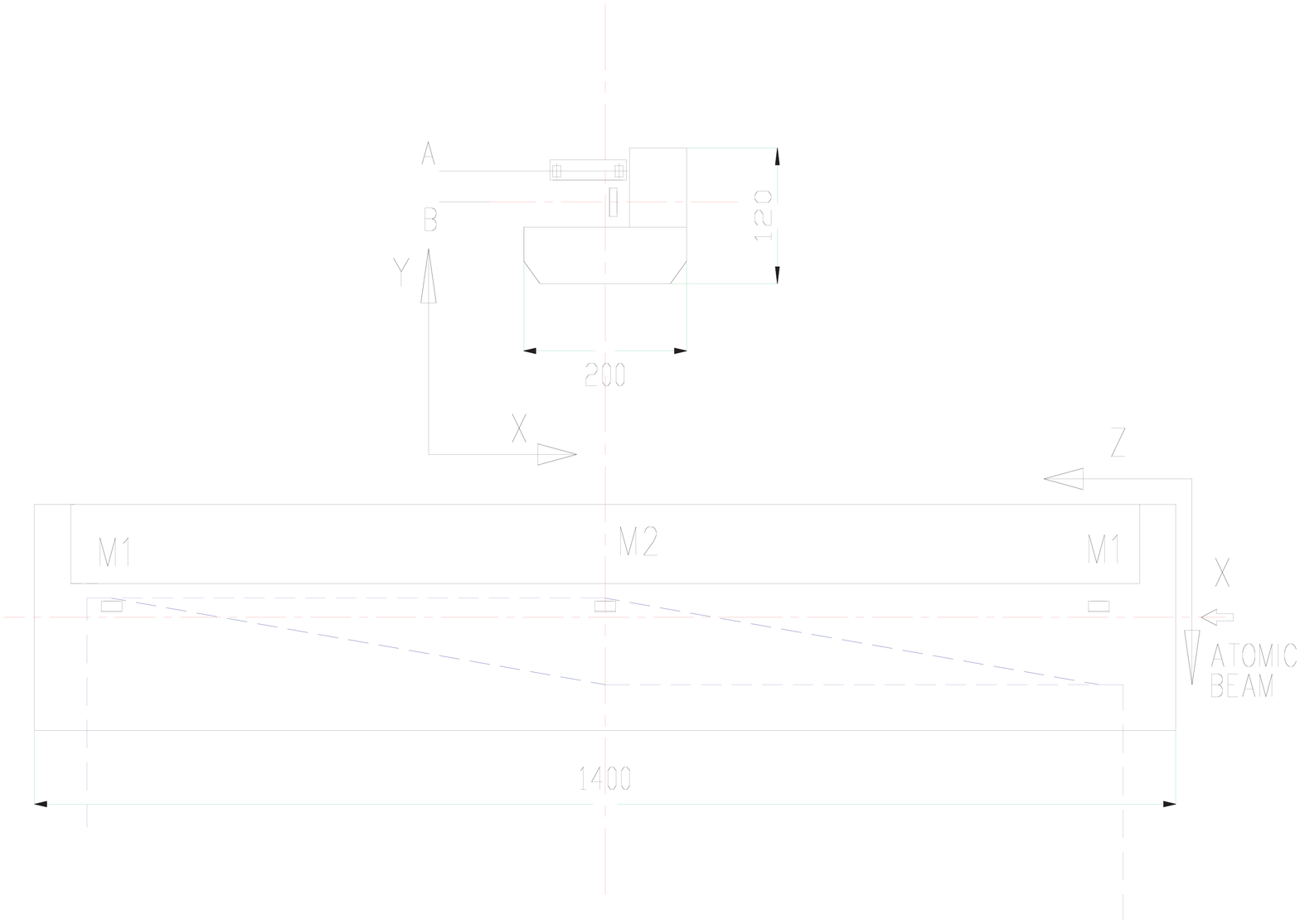}
\caption{\label{rail} Drawings of our interferometer rail showing
its shape and dimensions. Upper drawing: cross-section of the rail
showing the two blocks and their dimensions ($200\times 50$ mm$^2$
for the lower block; $70\times70$ mm$^2$ for the upper block). The
planes of the two interferometers are indicated (A for the atom
interferometer, B for the optical interferometer). Lower drawing:
top view of the rail, with some details: positions of the mirrors
$M_i$ for the three laser standing waves, position of the atomic
beam and of the laser beams of the Mach-Zehnder optical
interferometer.}
\end{figure}

%%%%%%%%%%%%%%%%%%%%%%%%%%%%%%%%%%%%%%%%%%%%%%%%%%%%%%

The rail will be described as an elastic object of length $2L$,
along the $z$ direction, which can bend only in the $x$ direction.
The rail is made of a material of density $\rho$ and Young's
modulus $E$. The cross-section, with a shape independent of the
$z$-coordinate, is characterized by its area $A= \int dx dy$ and
by the moment $I_y = \int x^2 dx dy$, the $x$-origin being taken
on the neutral line. The neutral line is described by a function
$X(z,t)$ which measures the position of this line with respect to
a Galilean frame linked to the laboratory (in this paper, we
forget that, because of Earth rotation, the laboratory is not a
Galilean frame). Elasticity theory \cite{landauelasticity67} gives
the equation relating the $t$- and $z$-derivatives of $X$:

\begin{equation}
\label{elasticity1} \rho A \frac{\partial^2 X}{\partial t^2} = -E
I_y \frac{\partial^4 X}{\partial z^4}
\end{equation}

\noindent The rail is submitted to forces and torques exerted by
its supports, which are related respectively to the third and
second derivatives of $X$ with respect to $z$:

\begin{equation}
\label{elasticity3} F_{x\epsilon} = - \epsilon E I_y
\frac{\partial^3 X}{\partial z^3}(z= \epsilon L)
\end{equation}

\begin{equation}
\label{elasticity4} M_{y\epsilon} =  \epsilon  E I_y
\frac{\partial^2 X}{\partial z^2}(z= \epsilon L)
\end{equation}

\noindent $\epsilon=\pm$ labels the rail ends at $z= \epsilon L$.
These torques and forces depend on the suspension of the rail. We
assume that the torques vanish, which would be exact if the
suspension was made in one point at each end and we consider that
the forces are the sum of an elastic term proportional to the
relative displacement and a damping term proportional to the
relative velocity:

\begin{equation}
\label{elasticity5} F_{x\epsilon} = - K_{\epsilon}\left[X(\epsilon
L,t) -x_{\epsilon}(t)\right] - \mu_{\epsilon}
\frac{\partial\left[X(\epsilon L,t) -x_{\epsilon}(t)
\right]}{\partial t}
\end{equation}

\noindent $x_{\epsilon}(t) $ is the coordinate of the support at
$z =\epsilon L$. The spring constants $K_{\epsilon}$ and the
damping coefficients $\mu_{\epsilon}$ may not be the same at the
two ends of the rail. The damping terms have an effective
character, because they represent all the damping effects.

\subsection{Solutions of these equations}

We introduce the Fourier transforms $X(z,\omega)$ and
$x_{\epsilon}(\omega)$ of the functions $X(z,t)$ and
$x_{\epsilon}(t)$. The general solution of equation
(\ref{elasticity1}) is:

\begin{equation}
\label{solution1} X(z,\omega) = a \sin(\kappa z) + b\cos(\kappa z)
+ c \sinh(\kappa z) + d\cosh(\kappa z)
\end{equation}

\noindent where $a$, $b$, $c$ and $d$ are the four
$\omega$-dependent amplitudes of the spatial components of the
function $X(z,\omega)$. $\omega$ and $\kappa$ are related by:

\begin{equation}
\label{solution2} \rho A \omega^2 = E I_y \kappa^4
\end{equation}

\noindent Equations (\ref{elasticity3}-\ref{elasticity5}) relate
$a, b, c, d$ to the source terms $x_{\epsilon}(\omega)$. Thanks to
the assumption $ M_{y\epsilon}=0$, $c$ and $d$ are related to $a$
and $b$:

\begin{eqnarray}
\label{solution3}
c & = & a \sin(\kappa L) / \sinh(\kappa L) \nonumber \\
d & =& b \cos(\kappa L) / \cosh(\kappa L)
\end{eqnarray}

\noindent and we get two equations relating $a$ and $b$ to
$x_{\epsilon}(\omega)$:

\begin{equation}
\label{solution4} \alpha_{\epsilon} a +  \epsilon \beta_{\epsilon}
b = \epsilon \gamma_{\epsilon} x_{\epsilon}(\omega)
\end{equation}

\noindent where $\alpha_{\epsilon}$, $\beta_{\epsilon}$ and
$\gamma_{\epsilon}$ are given in the appendix.

\subsection{Analysis of the various regimes}

To simplify, we assume that $K_{-} = K_{+} =K $ and $\mu_{-} =
\mu_{+} = 0$. $a(\omega)$ and $b(\omega)$ describe the transition
from a low-frequency dynamics in which the rail moves almost like
a solid to a high-frequency dynamics with a series of bending
resonances. When the frequency is low enough, $\kappa L \ll 1$
because $\kappa \propto \sqrt{\omega}$ is also small and we expand
the functions of $(\kappa L)$ up to third order (cubic terms in
$\kappa L$ are needed to transmit a transverse force through the
rail) and two resonances appear corresponding to pendular
oscillations of the rail. The first resonance appears on the $b$
amplitude, when $R$ given by equation (\ref{A5}) verifies $R
\approx 1$. This resonance corresponds to an in-phase oscillation
of the two ends of the rail, with a frequency $\omega_{osc} =
\sqrt{K/(\rho A L)}$. The second resonance, which appears on the
$a$ amplitude when $R\approx 3$, describes a rotational
oscillation of the rail around its center with a frequency
$\omega_{rot} = \omega_{osc} \sqrt{3}$. If the two spring
constants $K_{\epsilon}$ are different, these two resonances are
mixed (each resonance appears on the $a$ and $b$ amplitudes) and
their frequency difference increases.

For larger frequencies, $(\kappa L) $ is also larger and we cannot
use power expansions of the functions of $(\kappa L) $. We then
enter the range of bending resonances of the rail. If the forces
$F_{x\epsilon}$ are weak enough, these resonances are almost those
of the isolated rail which are obtained by writing that the
equation system (\ref{solution4}) has a nonvanishing solution when
the applied forces vanish and the resonance condition is:

\begin{equation}
\label{solution9} \cos(2\kappa L) \cosh(2\kappa L) =1
\end{equation}

\noindent which defines a series of $\kappa_n$ values given
approximately by:

\begin{equation}
\label{solution10} \kappa_n L \approx (2n +3)\frac{\pi}{4} +
\frac{(-1)^{n}}{\cosh\left[(2n+3)\pi/2\right]}
\end{equation}

\noindent $n$ starts from $0$ (a more accurate value of $\kappa_0
L $ is $\kappa_0 L = 2.365$) and $a= c=0$ when $n$ is even while
$b=d=0$ when $n$ is odd. $\omega_n$ is deduced from $\kappa_n$,
using equation (\ref{solution2}). For a given length $L$, the
wavevectors $\kappa_n$ are fixed, but the resonance frequencies
$\omega_n$ increase with the stiffness of the rail measured by the
quantity $EI_y/(\rho A)$. Finally, all the resonance frequencies
$\omega_n$ are related to $\omega_0$, by $\omega_n =\omega_0
(\kappa_n/\kappa_0)^2$ with:

\begin{equation}
\label{solution11} \omega_0 = 5.593 \sqrt{EI_y/(\rho A L^4)}
\end{equation}

\noindent Introducing the period $T_0$ of the first bending
resonance, $T_0 = 2\pi/\omega_0$, we may rewrite equation
(\ref{solution2}) in the form $ (\kappa L)^2 = 0.890 \times \omega
T_0$. Finally, we have calculated the $Q$ factors of the various
resonances (see appendix).

\subsection{Effect of vibrations on the interferometer signal}

The Fourier component $\Phi_p(\omega)$ of the phase $\Phi_p$ given
by equation (\ref{a5}) can be expressed as a function of the
amplitudes $a(\omega)$ and $b(\omega)$ given by solving the system
(\ref{solution4}). We assume that the gratings are on the neutral
line, which means that $x_i(t_i) = X(z_i,t_i)$ with $z_1 =
-L_{12}$ and $t_1=t -T$ for grating $G_1$, $z_2=0$ and $t_2 = t$
for grating $G_2$ and $z_3=+L_{12}$ and $t_3 = t+T$ for grating
$G_3$. We get:

\begin{eqnarray}
\label{excitation1} \frac{\Phi_p(\omega)}{2 p k_G} =  &  & \left[
b(\omega) \left(1 -\cos(\kappa L_{12})  + \left(1 -\cosh(\kappa
L_{12}) \right) \frac{\cos(\kappa L)}{\cosh(\kappa
L)}\right)\right. \nonumber \\ & & + \left. i a(\omega) \left(
\sin(\kappa L_{12}) + \sinh(\kappa L_{12}) \frac{\sin(\kappa
L)}{\sinh(\kappa L)} \right) \sin\left(\omega T \right) \right.
\nonumber \\ & & + \left. b(\omega) \left( \cos(\kappa L_{12})
+\cosh (\kappa L_{12}) \frac{\cos(\kappa L)}{\cosh(\kappa L)}
\right)  \left( 1 -\cos(\omega T)\right)\right]
\end{eqnarray}

\noindent where the different lines correspond to the bending, the
Sagnac and the acceleration terms in this order. We can simplify
this equation by making an expansion in powers of $(\omega T)$ up
to power $2$ and in powers of $\kappa L$ or $\kappa L_{12}$, up to
fourth order:

\begin{eqnarray} \label{excitation2} \frac{\Phi_p(\omega)}{p k_G} \approx
  &  &  b \left(  \frac{6(\kappa L)^2(\kappa
L_{12})^2-(\kappa L_{12})^4}{6}\right) \nonumber \\ & & + 4 i a
(\kappa L_{12}) \left(1- \frac{(\kappa L)^2}{6}\right) \left(
\omega T\right) \nonumber \\ & &  2 b (\omega T)^2
\end{eqnarray}

\noindent As in equation (\ref{a6}), we recognize the
instantaneous bending of the rail (first line, independent of the
time of flight $T$), the Sagnac term (second line, linear in $T$)
and the acceleration term (third line, proportional to $T^2$).
With the same approximations, $a$ and $b$ are given by equations
(\ref{A6},\ref{A7}). To further simplify the algebra, we replace
the distance $L_{12}$  by $L$ ($L_{12}$ will usually be close to
$L$) and we get:

\begin{eqnarray} \label{excitation4}
\frac{\Phi_p(\omega)}{p k_G} & \approx &  \left[x_{+} (\omega)
-x_{-}(\omega) \right] \frac{3i (\omega T)}{\left(3-R\right)}
 \nonumber
\\ & + & \left[x_{+}(\omega)  + x_{-}(\omega) \right]
 \frac{ 0.330 (\omega T_0)^2 + (\omega T)^2 }{2(1-R)}
\end{eqnarray}

\noindent where $R$ is given by equation (\ref{A5}).

These three equations (\ref{excitation1}, \ref{excitation2},
\ref{excitation4}) are the main theoretical results of the present
paper. Equation (\ref{excitation4}), which has a limited validity,
because of numerous approximations, gives a very clear view of the
various contributions. The first term, proportional to
$\left[x_{+} (\omega) -x_{-}(\omega)\right]$ and to the time of
flight $T$, describes the effect of the rotation of the rail
excited by the out of phase motion of its two ends. This term,
which is independent of the stiffness of the rail, is sensitive to
the rail suspension through the $(3-R)$ denominator. The second
term is the sum of the bending term, in $(\omega T_0)^2 $, and the
acceleration term, in $(\omega T)^2 $. Both terms are in
$\omega^2$ and they also have the same sensitivity to the
suspension of the rail, being sensitive to the first pendular
resonance, when $R\approx 1$. The bending term is small if the
rail is very stiff i.e. if the $T_0$ value is very small.

For larger frequencies ($\omega T \gtrsim 1$ or $\kappa L\gtrsim
1$), we must use numerical calculations of equation
(\ref{excitation1}). We may note that, because $\omega \gg
\omega_{osc} ,\omega_{rot} $, the $a$ and $b$ amplitudes will be
small excepted near the bending resonances which  appear either on
the $b$ or $a$ amplitudes and do not contribute to the same terms
of $\Phi_p(\omega)$.

\section{Application of the present analysis to our interferometer}

In this part, we are going to describe the rail of our
interferometer and to characterize its vibrations.

\subsection{Information coming from previous experiments}

When we built our interferometer in 1998, we knew that the
vibration amplitudes encountered by D. Pritchard
\cite{keith91,schmiedmayer97} and Siu Au Lee
\cite{giltner95b,giltner96} in their interferometers were large:
for instance $\sqrt{\left< \delta(t)^2\right>} \approx 500$ nm
after passive isolation by rubber pads in reference
\cite{keith91}. In this experiment, each grating was supported on
a flange of the vacuum pipe, which played the role of the rail. In
the interferometer of Siu Au Lee and co-workers, a rail inspired
by the three-rod design used for laser cavities was built
\cite{giltner96}, but with a rod diameter close to $15$ mm, the
rail was not very stiff. In both experiments, servo-loops were
used to reduce $\delta(t)$ to observe interference signals.

\subsection{The rail of our interferometer}

Rather than using servo-loops, we decided to achieve a very good
grating stability by building a very stiff rail. We had to choose
the material of the rail, its shape and its suspension, the main
constraint being that the rail had to fit inside the DN250 vacuum
pipe of our atom interferometer. The material must have a large
value of $E/ \rho$ ratio (Young's modulus divided by density): we
have chosen aluminium alloy rather than steel, both metals having
almost the same $E/ \rho$ ratio, because aluminium alloy is
lighter and easier to machine. The shape of the rail must give the
largest ratio $I_y/A$ with an open structure for vacuum
requirements: we choose to make the rail as large as possible in
the $x$ direction and rather thick to insure a good stiffness in
the $y$ direction, because the $x$ and $y$ vibrations are not
fully uncoupled. The rail, which is made of two blocks bolted
together, is represented in figure \ref{rail}. The lower block
($200 $ mm wide and $50$ mm thick) provides the rigidity. Its
length, $2L =1.4$ m, is slightly larger than twice the
inter-grating distance $L_{12} = 0.605$ m. The gratings, i.e. the
mirrors of the laser standing waves, are fixed to the upper block,
which has been almost completely cut in its middle to support the
central grating. As a consequence, its contribution to the
rigidity of the rail is probably very small and it will be
neglected in the following calculation of the first bending
resonance frequency $\omega_0/(2\pi)$: we use equation
(\ref{solution11}), with the full area $A\approx 1.49\times
10^{-2}$ m$^2$ but, for the moment $I_y$, we consider only the
lower block contribution ($I_y\approx 3.3 \times 10^{-5}$ m$^4$).
With $E= 72.4 \times 10^9$ N/m$^2$ and $\rho = 2.79 \times 10^3$
kg/m$^3$, we calculate $\omega_0/2 \pi \approx 437$ Hz.

When we built the suspension of the rail, the present analysis was
not available and we made a very simple suspension: the rail is
supported by three screws, two at one end and one at the other
end, so that it can be finely aligned. Each screw is supported on
a rubber block, model SC01 from Paulstra \cite{paulstra}. These
rubber blocks, made to support machine tools, are ring shaped with
a vertical axis. The technical data sheet gives only a rough
estimate of the force constant $K$ in the transverse direction, $K
\approx 10^6$ N/m. As the total mass of our rail $\rho A L \approx
58 $ kg, the pendular oscillations are expected to be at
$\omega_{osc}/(2\pi) \approx 20$ Hz and $\omega_{rot}/2\pi\approx
35$ Hz. We have not taken into account the mixing of these
resonances due to $K_- \neq K_+$, considering that the dominant
uncertainty comes from the spring constant values.

\subsection{Test of the vibrations by optical interferometry}

Following the works of the research groups of A. Zeilinger
\cite{gruber89,rasel95}, D. Pritchard
\cite{keith91,schmiedmayer97} and Siu Au Lee
\cite{giltner95b,giltner96}, the grating positions $x_i$  are
conveniently measured by a 3-grating Mach-Zehnder optical
interferometer. The phase $\Phi_{opt}$ of the signal of such an
optical interferometer is also given by equation (\ref{a6}), with
a negligible time delay $T$:

\begin{equation}
\label{exp2} \Phi_{opt} =  p k_{g,opt} \delta(t)
\end{equation}

\noindent We have built such an optical interferometer
\cite{miffre02}. The gratings from Paton Hawksley \cite{paton},
with 200 lines/mm ($k_{g,opt} = 3.14\times10^5$ m$^{-1}$), are
used in the first diffraction order with an helium-neon laser at a
$633$ nm wavelength. The excitation of the rail by the environment
gives very small signals, from which we deduce an upper limit of
$\sqrt{\left< \delta(t)^2\right>} <3$ nm. This result is close to
the noise (laser power noise and electronic noise) of the signal
and the noise spectrum has not revealed any interesting feature.

%%%%%%%%%%%%%%%%%%%%%%%%%%%%%%%%%%%%%%%%%%%%%%%%%%%%%%
\begin{figure}
\includegraphics[width = 8 cm,height= 6 cm]{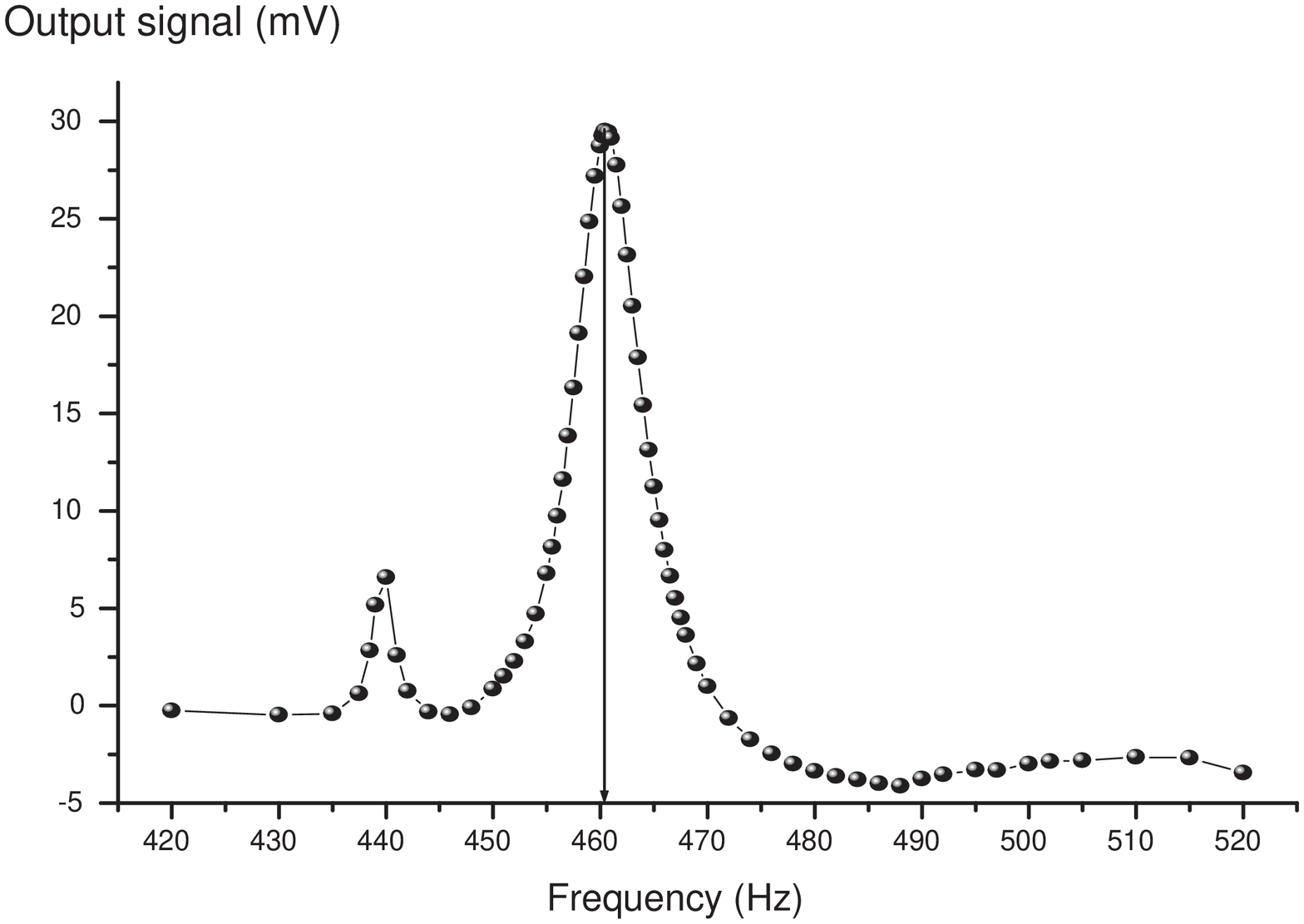}
\caption{\label{firstresonance} Response of the optical
interferometer to an excitation of the rail oscillation: the
modulation of the signal of the optical interferometer is plotted
as a function of the frequency of the sine wave sent to the
loudspeaker. We assign the main peak at $460.4$ Hz as due to the
first bending resonance but the weaker peak near $440$ Hz is not
assigned.}
\end{figure}

%%%%%%%%%%%%%%%%%%%%%%%%%%%%%%%%%%%%%%%%%%%%%%%%%%%%%%

Hence, we have made a spectroscopy of the rail vibrations in the
frequency domain by exciting its vibrations by a small loudspeaker
fixed on the rail, close to its center, with the coil moving in
the $x$-direction, so as to excite the $x$-bending of the rail.
The loudspeaker was excited by a sine wave of constant amplitude
and we have recorded with a phase-sensitive detection the
modulation of the optical interferometer signal. Figure
\ref{firstresonance} presents the detected signal in the region of
the first intense resonance centered at $\omega_0/(2\pi) = 460.4$
Hz, with a rather large Q-factor, $Q \approx 60$. We have also
observed a second resonance at $\omega_1/(2\pi) = 1375 $ Hz, with
$Q \approx 65 $, with a $30$ times weaker signal for the same
voltage applied on the loudspeaker (the $n=1$ resonance appears on
the $a$ amplitude and its detection by an optical interferometer,
sensitive only to the $b$ amplitude, is due to small asymmetries)
The first resonance frequency $460.4$ Hz is close to our estimate
$437$ Hz and the observed frequency ratio $\omega_1/\omega_0
\approx 2.99$ is also rather close to its theoretical value
$2.76$, so that we can assign these two resonances as the $n=0$
and $n=1$ bending resonances of the rail, the discrepancies being
due to oversimplifications of our model.

We have not observed any clear signature of the pendular
oscillations on the optical interferometer signal, probably
because the excitation and detection efficiencies are very low.
The detection of these pendular oscillations will be done in a
future experiment, using seismometers.

\subsection{Seismic noise spectrum: measurement and consequences for
the atom interferometer phase noise}

In the following calculation, we have not used our estimate of the
first pendular resonance $\omega_{osc}/(2\pi) \approx 20$ Hz,
because the predicted rms value of the bending $\sqrt{\left<
\delta(t)^2\right>}$ was considerably larger than measured. We
have used a larger value $\omega_{osc}/(2\pi) = 40$ Hz, with
$Q_{osc}= 16$ and the measured $\omega_0$ value, $\omega_0/(2\pi)
= 460.4$ Hz. In our model with the simplifying assumptions $K_{-}
= K_{+} =K $ and $\mu_{-} = \mu_{+} = \mu$, these three parameters
suffice to describe our rail and its suspension.

\begin{figure}
\includegraphics[width = 8 cm,height= 6 cm]{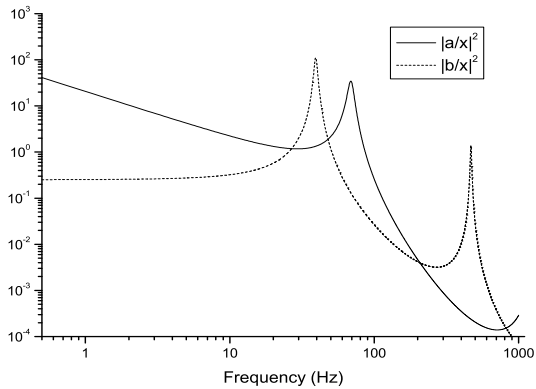}
\caption{\label{sensitivity} Logarithmic plot of the ratios
$\left|a(\omega)/x_{\epsilon}(\omega)\right|^2$ and
$\left|b(\omega)/x_{\epsilon}(\omega)\right|^2$ as a function of
the frequency $\nu= \omega/(2\pi)$. Three resonances appear, which
are the pendular oscillations, near $40$ and $69$ Hz, and the
first bending resonance near $460$ Hz.}
\end{figure}

In a first step, we calculate the $a$ and $b$ amplitudes as a
function of one noise amplitude $x_{\pm}(\omega)$, the other one
being taken equal to $0$. Figure \ref{sensitivity} plots the
ratios $\left|a(\omega)/x_{\epsilon}(\omega)\right|^2$ and
$\left|b(\omega)/x_{\epsilon}(\omega)\right|^2$ as a function of
the frequency $\nu= \omega/(2 \pi)$: three resonances appear in
the $1$-$10^3$ Hz range and, as expected, $a$ and $b$ decrease
rapidly when $\omega > \omega_{osc} ,\omega_{rot} $, a decrease
interrupted for $b$ by the first bending resonance.

%%%%%%%%%%%%%%%%%%%%%%%%%%%%%%%%%%%%%%%%%%%%%%%%%%%%%%

%%%%%%%%%%%%%%%%%%%%%%%%%%%%%%%%%%%%%%%%%%%%%%%%%%%%%%

\begin{figure}
\includegraphics[width = 8 cm,height= 6 cm]{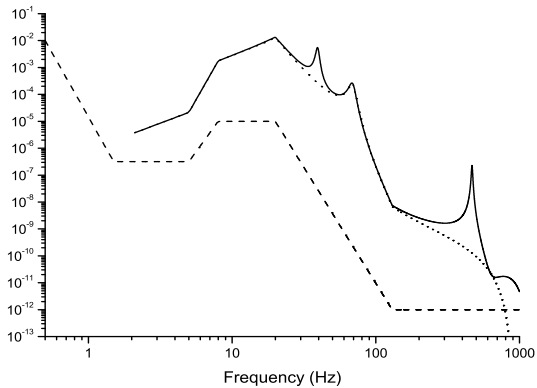}
\caption{\label{noisespectra} Calculated phase noise spectra
$\left|\Phi(\nu)/p\right|^2$ (full curve) and
$\left|\Phi_{Sagnac}(\nu)/p\right|^2$ (dotted curve), both in
rad$^2/$Hz as a function of the frequency $\nu$ in Hz. The
smoothed seismic noise spectrum $|x_{\epsilon}(\nu)|^2$ in
m$^2/$Hz used in the calculation is plotted (dashed curve) after
multiplication by $10^{10}$.}
\end{figure}

The seismic noise spectrum was recorded on our setup well before
the operation of our interferometer. This spectrum presents
several peaks appearing in the $8-60$ Hz range and most of these
peaks do not appear on a spectrum taken on the floor, because they
are due to resonances of the structure supporting the vacuum
pipes. As the peak frequencies have probably changed because of
modifications of the experiment since the recording, we have
replaced the recorded curve by a smooth curve just larger than the
measured spectrum. This noise spectrum $|x_{\epsilon}(\nu)|^2$ is
plotted in figure \ref{noisespectra}. We have also extended the
$\nu =0.5-100$ Hz frequency range to $\nu =0.5-10^3$ Hz, assuming
the noise to be constant when $10^2< \nu <10^3$ Hz.

Figure \ref{noisespectra} also plots the calculated phase noise
spectrum $\left|\Phi(\nu)/p\right|^2$, using equation
(\ref{excitation1}) and the Sagnac phase noise spectrum
$\left|\Phi_{Sagnac}(\nu)/p\right|^2$ deduced from equation
(\ref{excitation1}) by keeping only the term proportional to the
$a$ amplitude: clearly, the Sagnac phase noise is dominant except
near the in-phase pendular oscillation and the first bending
resonance. The bending resonance is in a region where the
excitation amplitude is very low, and, even after amplification by
the resonance $Q$-factor, the contribution of the bending
resonance to the total phase noise is fully negligible. In this
calculation, we have assumed that the two excitation terms
$x_{\epsilon}(\nu)$ have the same spectrum but no phase relation,
so that the cross-term $\left|x_{+1}(\nu)x_{-1}(\nu)\right|$ can
be neglected. This last assumption is bad for very low
frequencies, for which we expect $x_{+1}(\nu) \approx x_{-1}(\nu)$
(as the associated correction cancells the Sagnac term, we have
not extended the $\left|\Phi(\nu)/p\right|^2$ curves below $2$ Hz)
but this assumption is good as soon as the frequency is larger
than the lowest frequency of a resonance of the structure
supporting the vacuum chambers (near $8$Hz).

%%%%%%%%%%%%%%%%%%%%%%%%%%%%%%%%%%%%%%%%%%%%%%%%%%%%%%

%%%%%%%%%%%%%%%%%%%%%%%%%%%%%%%%%%%%%%%%%%%%%%%%%%%%%%

By integrating the phase noise over the frequency from $2$ up to
$10^3$ Hz, we get an estimate of the quadratic mean of the phase
noise:

\begin{equation}
\label{phasenoise1} \left<\Phi^2\right>  = 0. 16 p^2\mbox{ rad}^2
\end{equation}

\noindent This result is largely due to the Sagnac phase noise:
the same integration on the Sagnac phase noise gives
$\left<\Phi_{Sagnac}^2\right>  = 0. 13 p^2$ rad$^2$. We are going
to test this calculation, using the measurements of fringe
visibility as a function of the diffraction order $p$.

\subsection{Fringe visibility as a test of phase noise in atom interferometers}

A phase noise $ \Phi $ induces a strong reduction the fringe
visibility ${\mathcal{V}}$:

\begin{equation}
\label{d1} {\mathcal{V}}   = {\mathcal{V}}_{max} \exp\left[-
\left<\Phi^2 \right>/2 \right]
\end{equation}

\noindent assuming a Gaussian distribution of $\Phi$. When the
phase noise is due to inertial effects (see equation (\ref{a5})),
$\Phi$ is proportional to the diffraction order $p$, $\Phi_p = p
\Phi_1$. The fringe visibility $ {\mathcal{V}} $ is a Gaussian
function of the diffraction order $p$ \cite{delhuille02}:

\begin{equation}
\label{d2}{\mathcal{V}}  = {\mathcal{V}}_{max} \exp\left[-
\left<\Phi_1^2 \right> p^2 /2\right]
\end{equation}

\begin{figure}
\includegraphics[width = 8 cm,height= 6 cm]{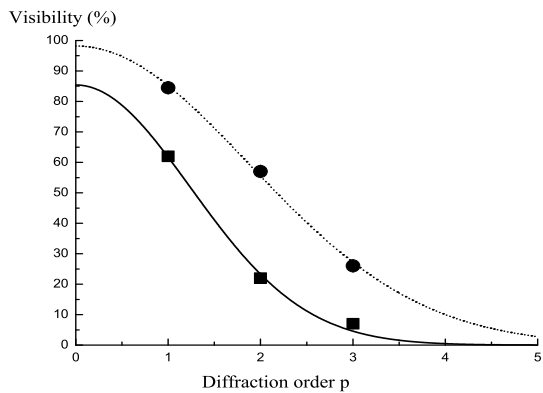}
\caption{\label{visibilityorder} Fringe visibility as a function
of the diffraction order $p$. Our measurements (round dots) are
fitted by equation (\ref{d2}) with ${\mathcal{V}}_{max} =98\pm 1$
\% and $\left<\Phi_1^2 \right> = 0.286 \pm 0.008$. The data points
of Giltner and Siu Au Lee (squares) are also fitted by equation
(\ref{d2}) with ${\mathcal{V}}_{max} =85\pm 2$ \% and
$\left<\Phi_1^2 \right> = 0.650 \pm 0.074$.}
\end{figure}

\noindent The atom interferometer of Siu Au Lee et al.
\cite{giltner95b,giltner96} and our interferometer \cite{miffre05}
have been operated with the first three diffraction orders. The
measured fringe visibility is plotted as a function of the
diffraction order in figure \ref{visibilityorder} and Gaussian
fits, following equation (\ref{d2}), represent very well the data.
The quality of these fits suggests that phase noise of inertial
origin is dominant and moreover that excellent visibility would be
achieved in the absence of phase noise. With our data points, we
deduce $\left<\Phi_p^2 \right> = (0. 286 \pm 0.008) p^2$. Our
estimate given by equation (\ref{phasenoise1}) is $56$\% of this
value and, considering the large uncertainty on several parameters
(seismic noise, frequency and $Q$ factors of the pendular
resonances), the agreement can be considered as good.

%%%%%%%%%%%%%%%%%%%%%%%%%%%%%%%%%%%%%%%%%%%%%%%%%%%%%%

%%%%%%%%%%%%%%%%%%%%%%%%%%%%%%%%%%%%%%%%%%%%%%%%%%%%%%

\section{How to further reduce the vibration phase noise in
3-grating Mach-Zehnder atom interferometers.}

The phase noise induced by vibrations is very important and its
reduction will considerably improve the operation of atom
interferometers.

\subsection{Servoloops on the grating positions}

Pritchard and co-workers \cite{keith91,schmiedmayer97} as well as
Giltner and Siu Au Lee \cite{giltner95b} have used servo-loops to
reduce the vibrational motion of the grating. The error signal was
given by the optical Mach-Zehnder interferometer, which measures
the instantaneous bending $ \delta(t) = \left(2 x_2(t) - x_1(t)
-x_3(t) \right)$ and, as recalled above, in both experiments, the
error signal before correction was large. In the experiment of
Pritchard and co-workers, the correction was applied to the second
grating. In the limit of a perfect correction, the bending term in
equation (\ref{a6}) is cancelled and this correction does not
modify the Sagnac and the acceleration terms. The fact that acting
on the second grating has no inertial effects is a somewhat
surprizing result, which can be explained by the symmetry of the
Mach-Zehnder interferometer. In the experiment of Giltner and Siu
Au Lee, the correction, which was applied to the third grating,
cancels $\delta(t)$ but the Sagnac and acceleration terms are
enhanced. In any case, the servo-loop can reduce the instantaneous
bending $\delta(t)$ but it cannot reduce the Sagnac and
acceleration terms. We think that a very stiff rail is a better
solution for earth-based interferometers. For space based
experiments, the phase noise spectra due to inertial vibration is
different and the above solution may not be optimum, because of
the large weight of the rail.

\subsection{Possible improvements of the rail}

The stiffness of our rail has reduced to a low level the bending
and acceleration terms in the phase-noise of our interferometer.
In our model, the rail stiffness is measured by only one
parameter, the period $T_0$ of the lowest bending resonance, which
scales with the rail length $L$ like $L^2$. Our $T_0$ value, $T_0
= 2.2 \times 10^{-3}$ s, is still $3.8$ times larger than the time
flight $T \approx 5.7 \times 10^{-4}$ s in our experiment (lithium
beam mean velocity $u = 1065$ m/s; inter-grating distance $L_{12}
= 0.605$ m) and the bending term in equation (\ref{excitation2})
is $3$ times larger than the acceleration term. We can further
reduce the bending term by reducing $T_0$, either by using an
I-shaped rail to increase the $I_y/A$ ratio or by using a material
with a larger $E/\rho$ ratio than aluminium alloy (for example,
silicon carbide).

A defect of our rail is that it has no symmetry axis and the $x$
and $y$ bending modes are partly mixed. As the moment $I_x$ is
considerably smaller than $I_y$, the bending resonances in the
$y$-direction are at lower frequencies than in the $x$-direction.
A better rail design should decouple almost completely the $x$ and
$y$ vibrations.

\subsection{Possible improvements of the suspension}

The suspension of our rail is very primitive, with rather large
spring constants and pendular resonances probably in the $\nu=
20-100$ Hz range. A very different choice was made by J. P.
Toennies and co-workers \cite{toennies03}: the rail was suspended
by wires, the restoring forces being due to gravity. The pendular
oscillation frequency is $\omega_{osc} = \sqrt{g/l}$, where $l$ is
the wire length. For a typical $l$ value, $l =10$ cm,
$\omega_{osc}/(2 \pi ) \approx 1.5 $ Hz. In this experiment, a
servo-loop was necessary to reduce the amplitudes of the pendular
motions.

From the seismic noise spectrum of figure \ref{noisespectra}, it
seems clear that the resonances of the suspension should not be in
the $5$ to $30$ Hz range, where there is an excess noise. Our
choice is not ideal and the choice of J. P. Toennies and
co-workers \cite{toennies03} seems better, as the seismic noise in
the $8-30$ Hz range can be largely reduced. Lower pendular
resonance frequencies can be achieved by clever design (crossed
wire pendulum, Roberts linkage) and a large know-how has been
developed for the construction of gravitational wave detectors
LIGO, VIRGO, GEO, TAMA, etc. Without aiming at a comparable level
of performance, it should be possible to build a very efficient
suspension.

\subsection{Fringe visibility in atom interferometers}

\begin{figure}
\includegraphics[width = 8 cm,height= 6 cm]{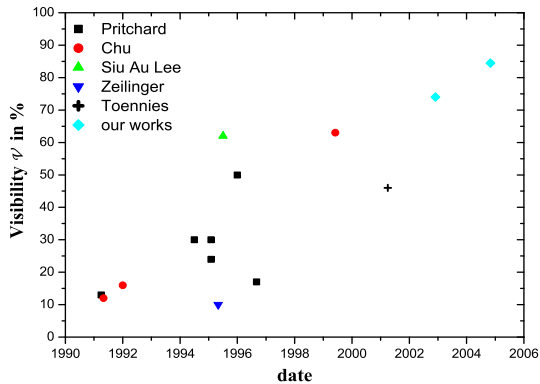}
\caption{\label{visibilitytime} Fringe visibility ${\mathcal{V}} $
in three-grating atom interferometers as a function of the date of
submission. These data points are taken from the following
publications: Pritchard's group
\cite{keith91,ekstrom95,schmiedmayer95,roberts02,lenef97,chapman95,
schmiedmayer97}; Chu's group
\cite{kasevich91,kasevich92,peters99}; Siu Au Lee's group
\cite{giltner95b,giltner96}; Zeilinger's group \cite{rasel95};
Toennies's group \cite{meschede03}; our works
\cite{delhuille02a,miffre05}.}
\end{figure}

Since the first atom interferometry experiments in 1991, many
different interferometers have been operated and numerous efforts
have been done to improve these experiments. We are going to
review the achieved fringe visibility, as this quantity is very
sensitive to phase noise and other phase averaging effects
(wavefront distorsions, $M$-dependent phase due to magnetic field
gradient, etc). We have considered only Mach-Zehnder atom
interferometers in which the atom paths are substantially
different, excluding for instance atomic clocks. Our review is not
complete, in particular because some publications do not give the
fringe visibility. The measured values of the visibility are
plotted in figure \ref{visibilitytime}. Some low values are not
only due to phase noise but also to other reasons: maximum
visibility less than $100$ \% in the case of Moiré detection
\cite{keith91}, parameters chosen to optimize the phase
sensitivity \cite{lenef97}. Over a $15$-year period, impressive
progress have been achieved and, hopefully, the same trend will
continue in the future. The comparison with optical interferometry
is encouraging as very high fringe visibility is routinely
achieved in this domain.

%%%%%%%%%%%%%%%%%%%%%%%%%%%%%%%%%%%%%%%%%%%%%%%%%%%%%%

%%%%%%%%%%%%%%%%%%%%%%%%%%%%%%%%%%%%%%%%%%%%%%%%%%%%%%

\section{Conclusion}

The  present paper has analyzed the phase noise induced in a
Mach-Zehnder atom interferometer by mechanical vibrations. We have
first recalled the inertial sensitivity of atom interferometers,
following the presentation of Schmiedmayer et al.
\cite{schmiedmayer97}. We have developed a simple $1$D model of
the rail supporting the diffraction gratings. This model gives an
unified description of the low-frequency dynamics, in which the
rail behaves as a solid object, and the high frequency domain, in
which rail bending cannot be neglected.

We have then described the rail of our interferometer. Our design
has produced a very stiff rail and the bending of the rail due to
vibrations appears to be almost negligible, while it was important
in several previous experiments. In the low-frequency range, up to
the frequency of the rotational resonance of the rail suspension,
the out-of-phase vibrations of the two ends of the rail induce
rotations of the rail, which are converted in phase noise by
Sagnac effect: this is the dominant cause of inertial phase noise
in our interferometer. A rapid decrease of the fringe visibility
with the diffraction order has been observed by Siu Au Lee and
co-workers \cite{giltner95b,giltner96} and by our group
\cite{miffre05}: the observed behavior is well explained as due to
an inertial phase-noise and the deduced phase noise value is in
good agreement with a value deduced from our model of the rail
dynamics, using as an input the seismic noise measured on our
setup.

In the last part, we have presented a general discussion of the
vibration induced phase noise in 3-grating Mach-Zehnder
interferometers. A reduction of this noise is absolutely necessary
in order to operate atom interferometers either with higher
diffraction orders or with slower atoms. In our experiment, a
large reduction of this noise can be obtained by improving the
suspension of the interferometer rail. Finally, we have reviewed
the published values of the fringe visibility obtained with atom
interferometers, thus illustrating the rapid progress since 1991.

\section{Acknowledgements}

We have received the support of CNRS MIPPU, of ANR and of R\'egion
Midi Pyr\'en\'ees through a PACA-MIP network. We thank A. Souriau
and J-M. Fels for measuring the seismic noise in our laboratory.

\section{Appendix: amplitudes of vibration of the rail and $Q$ factors of its resonances}

Equations (\ref{elasticity3}) and (\ref{elasticity5}) relate the
values of the $a$,  $b$, $c$, $d$ amplitudes to
$x_{\epsilon}(\omega)$. Using equation (\ref{solution3}), we
eliminate $c$ and $d$ to get the system of equations
(\ref{solution4}) with:

\begin{eqnarray}
\label{A2} \alpha_{\epsilon} &=& \left[\cosh(\kappa L)\sin(\kappa
L) - \sinh(\kappa L)\cos(\kappa L)\right]\cosh(\kappa L)\nonumber
\\ & - & 2(\kappa L)\cosh(\kappa L) \sinh(\kappa L) \sin(\kappa
L)R_{\epsilon}^{-1} \nonumber
\\ \beta_{\epsilon} &=& \left[\cosh(\kappa L)\sin(\kappa L) +
\sinh(\kappa L)\cos(\kappa L)\right] \sinh(\kappa L)\nonumber \\
&-& 2(\kappa L) \cosh(\kappa L) \sinh(\kappa L)\cos(\kappa
L)R_{\epsilon}^{-1} \nonumber \\ \gamma_{\epsilon}& =& - (\kappa
L)\cosh(\kappa L) \sinh(\kappa L)R_{\epsilon}^{-1}
\end{eqnarray}

\noindent with $R_{\epsilon}= \rho A L\omega^2/\left(K_{\epsilon}
- i \mu_{\epsilon} \omega\right)$. From now on, $K_{-1} = K_{+1}
=K $ and $\mu_{-1} = \mu_{+1} = \mu$. Then $\alpha$, $\beta$ and
$\gamma$ are independent of $\epsilon$. $R$ can be expressed as a
function of $\omega_{osc}= \sqrt{K/(\rho A L)}$ and $Q_{osc}=
(\rho A L \omega_{osc})/\mu$:

\begin{equation}
\label{A5} R= \omega^2 /\left[\omega_{osc}^2
-i\frac{\omega_{osc}\omega}{ Q_{osc}}\right]
\end{equation}

\noindent We get $a$ and $b$:

\begin{eqnarray}
\label{A4} a &=& \gamma(x_+ - x_-)/(2\alpha) \nonumber \\ b &=&
\gamma(x_+ + x_-)/(2\beta)
\end{eqnarray}

\noindent When $\kappa L \ll 1$, by expanding $\alpha$, $\beta$
and $\gamma$ in power of $\kappa L$ (up to the third order for
$\alpha$), we get:

\begin{equation} \label{A6}
 a  = \frac{x_+ -x_-}{4 \kappa L} \times \frac{3}{3-R}
\end{equation}
\begin{equation} \label{A7}
 b = \frac{x_+ + x_-}{4(1-R)}
\end{equation}
\noindent $b$ exhibits a resonance when $R=1$ ($\omega
=\omega_{osc}$) and $a$ when $R=3$ ($\omega =\omega_{osc}
\sqrt{3}$). We have calculated the resonance $Q$ factors, in the
weak damping limit. For an isolated resonance, the $Q$ factor is
related by $Q= 2 \pi E_{tot}/\Delta E$ to the total energy
$E_{tot}$ and the energy $\Delta E$ dissipated during one
vibration period. We get:

\begin{equation}
\label{B2} Q_{osc}= \rho A L  \omega_{osc}/\mu
\end{equation}

\begin{equation}
\label{B3} Q_{rot}= \rho A L  \omega_{rot}/(3 \mu)
\end{equation}

\begin{equation}
\label{B4} Q_n= \rho A L \omega_n g(\kappa_n L) /(8\mu)
\end{equation}

\noindent where the function $g(\kappa_n L)$ depends on the parity
of $n$:
\begin{eqnarray}
\label{B5} g(\kappa_n L) & = &  \left[1 + \frac{\sin(2\kappa_n
L)}{2\kappa_n L}\right] \left[\frac{1}{\cos^2(\kappa_n L)}  +
\frac{1}{\cosh^2(\kappa_n L)} \right] \mbox{ for even } n
\nonumber \\ & = & \left[1 - \frac{\sin(2\kappa_n L)}{2\kappa_n
L}\right] \left[\frac{1}{\sin^2(\kappa_n L)}  +
\frac{1}{\sinh^2(\kappa_n L)} \right] \mbox{ for odd } n
\end{eqnarray}

\noindent From the measured $Q$-factor of the first bending
resonance ($n=0$), we get $\mu \approx 560$ kg.s$^{-1}$.

%%%%%%%%%%%%%%%%%%%%%%%%%%%%%%%%%%%%%%%%%%


\begin{references}

\bibitem {anandan77} J. Anandan, Phys. Rev. D {\bf 15}, 1448 (1977)

\bibitem {clauser88} J. F. Clauser, Physica B {\bf 151}, 262 (1988)

\bibitem {kasevich91} M. Kasevich and S. Chu, Phys. Rev. Lett. {\bf 67}, 181 (1991)

\bibitem {kasevich92} M. Kasevich and S. Chu, Appl. Phys. B {\bf 54}, 321 (1992)

\bibitem {cahn97} S. B. Cahn, A. Kumarakrishnan, U. Shim, T.
Sleator, P. R. Berman and B. Dubetsky, Phys. Rev. Lett. {\bf 79},
784 (1997)

\bibitem {peters99} A. Peters, K. Y. Chung and S. Chu, Nature {\bf 400}, 849 (1999)

\bibitem {peters01} A. Peters, K. Y. Chung and S. Chu, Metrologia {\bf 38}, 25 (2001)

\bibitem {snadden98} M. J. Snadden, J. M. McGuirk, P. Bouyer, K. G. Haritos and M. A.
Kasevich , Phys. Rev. Lett. {\bf 81}, 971 (1998)

\bibitem {mcguirk02} J. M. McGuirk, G. T. Foster, J. B. Fixler, M. J. Snadden, and M.
A. Kasevich, Phys. Rev. A {\bf 65}, 033608 (2002)

\bibitem{tino02} G. M. Tino, Nucl. Phys. B {\bf 113}, 289 (2002)

\bibitem{riehle91} F. Riehle, Th. Kisters, A. Witte, J. Helmcke and Ch. J. Bordé,
Phys. Rev. Lett. {\bf 67}, 177 (1991)

\bibitem{lenef97} A. Lenef, T. D. Hammond, E. T. Smith, M. S. Chapman, R. A.
Rubenstein, and D. E. Pritchard, Phys. Rev. Lett. {\bf 78}, 760
(1997)

\bibitem{gustavson97} T. L. Gustavson, P. Bouyer and M. A. Kasevich,
Phys. Rev. Lett. {\bf 78}, 2046 (1997)

\bibitem{gustavson00} T. L. Gustavson, A. Landragin and M. A. Kasevich,
Class. quantum Grav. {\bf 17}, 2385 (2000)

\bibitem{leduc04} F. Leduc, D. Holleville, J. Fils, A. Clairon, N. Dimarcq, A. Landragin,
P. Bouyer and Ch. J. Bord\'e, Proceedings of 16th ICOLS, P.
Hannaford et al. editors, World Scientific (2004)

\bibitem{landragin04} A. Landragin et al., Proceedings of ICATPP-7, World Scientific (2002)

\bibitem{keith91} D. W. Keith, C. R. Ekstrom, Q. A. Turchette and
D. E. Pritchard, Phys. Rev. Lett. {\bf 66}, 2693 (1991)

\bibitem{schmiedmayer97} J. Schmiedmayer, M. S. Chapman, C. R.Ekstrom,
T. D. Hammond, D. A. Kokorowski, A. Lenef, R.A. Rubinstein, E. T. Smith
and D. E. Pritchard, in Atom interferometry edited by P. R. Berman (Academic Press 1997), p 1

\bibitem{turchette92} Q. Turchette, D. Pritchard and D. Keith, J. Opt. Soc. Am. B {\bf9}, 1601 (1992)

\bibitem{champenois99} C. Champenois, M. B\"uchner and J. Vigu\'e, Eur. Phys. J. D {\bf 5}, 363 (1999)

\bibitem{landauelasticity67} L. Landau and E. Lifchitz, Theory of Elasticity, Pergamon Press,
Oxford (1986)

\bibitem{paulstra} Paulstra company, website http://www.paulstra-vibrachoc.com

\bibitem{gruber89} M. Gruber, K. Eder and A. Zeilinger, Phys.
Lett. A {\bf 140}, 363 (1989)

\bibitem{rasel95} E. M. Rasel, M. K. Oberthaler, H. Batelaan,
J. Schmiedmayer and A. Zeilinger, Phys. Rev. Lett., {\bf 75}, 2633
(1995)

\bibitem{giltner95b} D.M. Giltner, R. W. McGowan and Siu Au Lee,
Phys. Rev. Lett., {\bf 75}, 2638 (1995)

\bibitem{giltner96} D. M. Giltner, Ph. D. thesis, Colorado State University, Fort Collins (1996)

\bibitem{miffre02} A. Miffre, R. Delhuille, B. Viaris de Lesegno, M. B\"uchner,
C. Rizzo and J. Vigu\'e, Eur. J. Phys. {\bf 23}, 623 (2002)

\bibitem{paton} Paton Hawksley Education Ltd, UK, website: http://www.patonhawksley.co.uk/

\bibitem{toennies03} J. P. Toennies, private communication (2003)

\bibitem{delhuille02} R. Delhuille, A. Miffre, B. Viaris de Lesegno, M. B\"uchner,
C. Rizzo, G. Tr\'nec and J. Vigu\'e, Acta Physica Polonica {\bf
33}, 2157 (2002)

\bibitem{miffre05} A. Miffre, M. Jacquey, M. B\"uchner, G. Tr\'enec
and J. Vigu\'e, Eur. Phys. J. D {\bf 33}, 99 (2005)

\bibitem{ekstrom95} C. R. Ekstrom, J. Schmiedmayer, M. S. Chapman,
T. D. Hammond and D. E. Pritchard, Phys. Rev. A {\bf 51}, 3883
(1995)

\bibitem{schmiedmayer95} J. Schmiedmayer, M. S. Chapman, C. R.
Ekstrom, T. D. Hammond, S. Wehinger and D. E. Pritchard, Phys.
Rev. Lett. {\bf 74}, 1043 (1995)

\bibitem{roberts02} T. D. Roberts, A. D. Cronin, D. A. Kokorowski1,
and D. E. Pritchard, Phys. Rev. Lett. {\bf 89}, 200406 (2002))

\bibitem{chapman95} M. S. Chapman, C. R. Ekstrom, T. D. Hammond,
R. A. Rubenstein, J. Schmiedmayer, S. Wehinger, and D. E.
Pritchard, Phys. Rev. Lett. {\bf 74}, 4783 (1995)

\bibitem{meschede03} work of J. P. Toennies and R. Br\"uhl quoted
in D. Meschede, Gerthsen Physik, {\bf 22}, 709 (2003)

\bibitem{delhuille02a} R. Delhuille, C. Champenois, M. B\"uchner, L.
Jozefowski, C. Rizzo, G. Tr\'enec and J. Vigu\'e, Appl. Phys. {\bf
B 74}, 489 (2002)


\end{references}
\end{document}